\documentclass[electronic, preprint]{vgtc}   

\graphicspath{{figures/}} % where to search for the images

\usepackage{times}                     % we use Times as the main font
\usepackage{tabu}                      % only used for the table example
\usepackage{booktabs}                  % only used for the table example
\usepackage{lipsum}                    % used to generate placeholder text
\usepackage{mwe}                       % used to generate placeholder figures

\usepackage{mathptmx}                  % use matching math font

\usepackage[dvipsnames,svgnames,xcdraw, table]{xcolor}
\PassOptionsToPackage{pagebackref,bookmarks}{hyperref}
\usepackage{orcidlink}
\usepackage{cite}
\usepackage{flushend}

\usepackage{balance}

\onlineid{0}

\vgtccategory{Research}

\vgtcinsertpkg

\preprinttext{This article was accepted to the 2024 \href{https://firstpersonvis.github.io/}{First-Person Visualizations for Outdoor Physical Activities Workshop} held as part of \href{http://ieeevis.org/year/2024/welcome}{IEEE VIS 2024}}

\title{Collecting Information Needs for Egocentric Visualizations while Running}

\author{Ahmed Elshabasi\thanks{e-mail: ahmed.elshabasi@ucalgary.ca\orcidlink{}}\\ %
     \parbox{1.8in}{\scriptsize \centering University of Calgary}%
\and Lijie Yao\thanks{e-mail: yaolijie0219@gmail.com \orcidlink{0000-0002-4208-5140}}\\ %
     \parbox{1.8in}{\scriptsize \centering Xi'an Jiaotong-Liverpool University \\ University of Calgary \\ Universit{\'e} Paris-Saclay, CNRS, Inria, LISN, France}%
\and Petra Isenberg\thanks{e-mail: petra.isenberg@inria.fr \orcidlink{0000-0002-2948-6417}}\\ %
     \parbox{1.8in}{\scriptsize \centering Universit{\'e} Paris-Saclay CNRS, Inria, LISN, France}%
\and Charles Perin\thanks{e-mail: cperin@uvic.ca \orcidlink{0000-0002-7324-9363}}\\ %
     \parbox{1.8in}{\scriptsize \centering University of Victoria}%
\and Wesley Willett\thanks{e-mail:  wesley.willett@ucalgary.ca \orcidlink{0000-0002-6793-3062}}\\ %
     \parbox{1.8in}{\scriptsize \centering University of Calgary}%
}

\teaser{
  \centering
  \includegraphics[width=\linewidth]{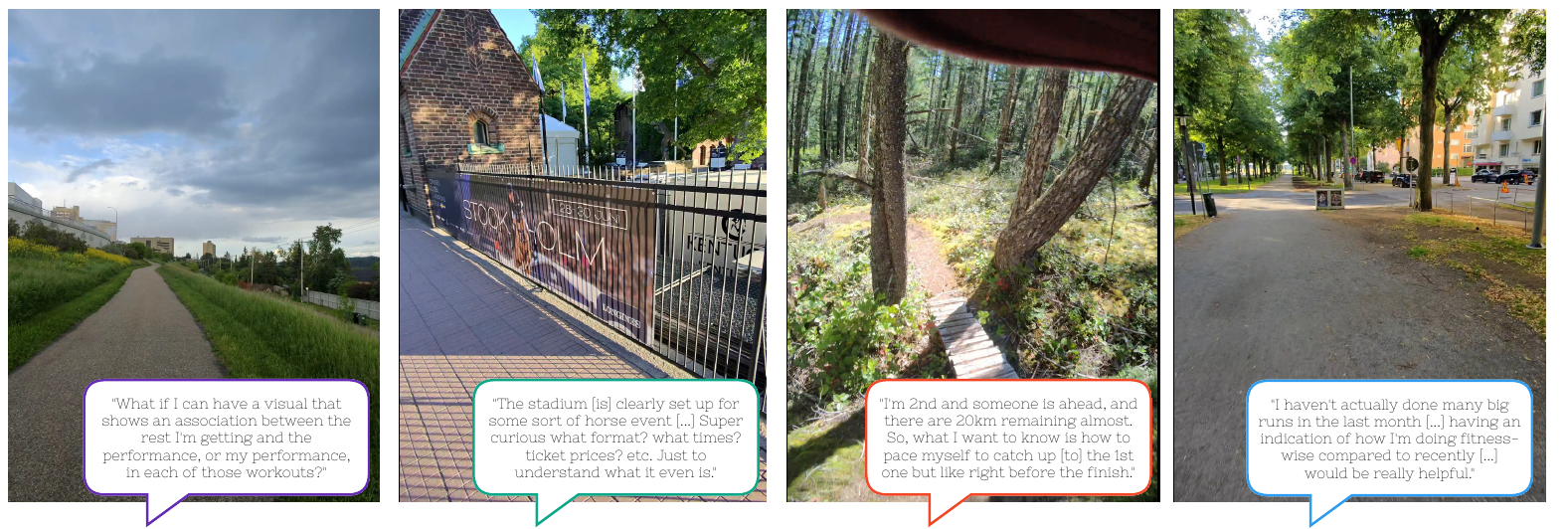}
  \caption{Four information needs, plus accompanying video stills captured during real-world runs and races highlight opportunities for new egocentric visualizations.}
  \label{fig:teaser}
}

\abstract{
    We investigate research challenges and opportunities for visualization in motion during outdoor physical activities via an initial corpus of real-world recordings that pair egocentric video, biometrics, and think-aloud observations. %
    With the increasing use of tracking and recording devices, such as smartwatches and head-mounted displays, more and more data are available in real-time about a person's activity and the context of the activity. 
    However, not all data will be relevant all the time. Instead, athletes have information needs that change throughout their activity depending on the context and their performance. 
    To address this challenge, we describe the collection of a diverse corpus of information needsontextualizing audio, video, and sensor data. Next, we propose a first set of research challenges and design considerations that explore the difficulties of visualizing those real data needs in-context and demonstrate a prototype tool for browsing, aggregating, and analyzing this information. Our ultimate goal is to understand and support embedding visualizations into outdoor contexts with changing environments and varying data needs.  
} 

\begin{document}

\firstsection{Introduction}
\maketitle

The integration of situated visualizations into sports and physical activities presents an exciting frontier for enhancing athletic performance and experience. With the proliferation of wearable technology and real-time data collection, athletes have unprecedented access to detailed information about their activities. Such information can offer athletes immediate, actionable insights that can inform their training and performance. However, the challenge lies in effectively identifying the athletes' data needs and presenting this information in a manner that is contextually relevant and dynamically responsive to their changing needs and environments. 

Our work presents an initial corpus of contexts and information needs captured live by amateur athletes during outdoor physical activities (primarily running). The corpus includes biometric data, egocentric videos captured by the athletes, think-aloud audio documenting their in-context information needs, and written transcripts of those needs. Additionally, we highlight early progress on a tool for analyzing these needs and their surrounding context and supporting the design of new visualizations to address them.
Our initial analysis of information needs from the corpus reveal considerable variability in needs across individuals, even for similar activities, along with a number of other potential challenges and opportunities for designing immersive visualization-in-motion applications.

\section{Related Work}
Our work is closely related to the topic of situated and embedded visualizations~\cite{Willett:2017:EmbeddedDataRepresentation}, which are visual representations located close to a data referent~\cite{Bressa:2019:SketchingIdeationforSituatedVisDesign}. 
In addition, we are inspired by the research direction of visualization in motion in which visualizations are used during relative movement between a viewer and a visualization \cite{Yao:2022:VisinMotion}. 
We also build on visual analytics research for outdoor physical activities, focusing on the connection between situated data needs and their contexts under dynamic situations and how to visualize such needs best. 
    
Physical activities in dynamic outdoor scenarios have gained a lot of traction in the visualization community. The research ranges from analyzing the data outside of its context~\cite{Perin:2013:SoccerAnalysis, Perin:2014:ATable, Perin:2016:GapChart, Perin:2018:StateSport, Wang:2021:TableTenis, Wu:2018:TableTennis, Andrienko:2021:FootballVis, Pingali:2001:BadmintonAnalysis} to embedding visual representations into scenarios, such as swimming~\cite{Yao:2024:VisForSwimming}, table tennis~\cite{Chen:2023:Sporthesia, Chen:2021:VisCommentator}, soccer~\cite{Wu:2018:TableTennis, Manuel:2017:SoccerTeamSportAnalysis}, basketball~\cite{Lin:2023:BallCourt, Chen:2023:iBall, Lin:2022:Omnioculars, Lin:2021:ARVisforBasketballTraining, Bai:2016:BasketballAnalysis}, cycling \cite{Kaplan:2016:CyclingForcesAnalysis}, badminton~\cite{Lin:2024:VIRD, Chu:2022:TIVEEBadminton, Ye:2021:BadmintonVis3D}, and tennis~\cite{Polk:2020:TenisMatchVis}. These works mainly focus on providing technologies and tools that support injecting visual representations into various scenarios. 
    
Only few past works elicited viewers' actual data needs~\cite{Yao:2024:VisForSwimming,Lin:2022:Omnioculars}.  Yao et al.~\cite{Yao:2024:VisForSwimming} investigated the audiences' real data interests when watching swimming races and provided a technology probe, \textit{SwimFlow}, that allows viewers to embed visualizations that can move with swimmers in real-time into swimming videos according to their real data needs. 
Lin et al.~\cite{Lin:2022:Omnioculars} conducted a survey with basketball fans to understand  important moments in basketball games and provided concrete visualization solutions for these spot moments. 
Our work is also based on real data needs but it is unique in that, instead of considering the audience, we focus on the athletes themselves and how their needs change over time, based on their performance, and the context of their activity.

Other works have employed various approaches in analyzing running data for prospective design applications. A study on designing for active environments \cite{van2024changing} used data from running apps to explore urban design. Through the collective lens, the analysis revealed general running patterns and environmental preferences, while the individual lens focused on unique user behaviors. Contextual data such as weather, neighborhood info, and lighting conditions were also considered to understand environmental influences. A similar study examined how environmental characteristics and runners' motives affect their perception of running environments \cite{deelen2019attractive}, collecting data from half marathon runners through the Eindhoven Running Survey 2015. The researchers used linear regression analyses to assess the impact of intrapersonal characteristics and perceived environmental traits. 
    
\section{Preparation: Collecting Information Needs}

\begin{figure}
    \centering
    \includegraphics[width=0.9\linewidth]{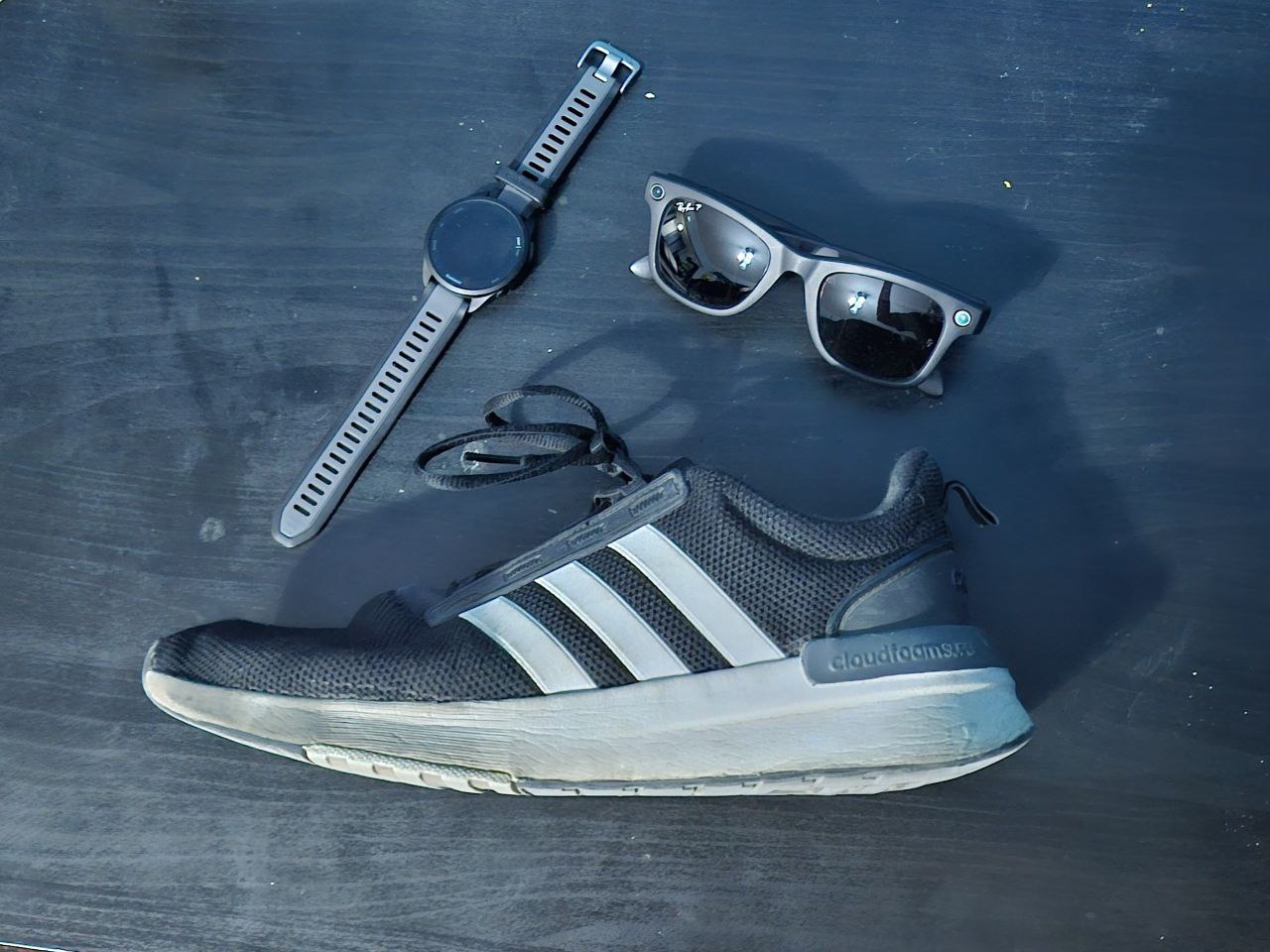}
    \caption{Our lightweight workout equipment for capturing runs, information needs, and biometric data: a Ray-Ban Meta Smart Glasses and a Garmin smartwatch.}
    \label{fig:enter-label}
\end{figure}

To study different in-situ data needs in various contexts of outdoor physical activities, we are currently working on a database of first person videos of these activities. The database is meant to collect short video snippets together with audio recordings of information needs.
We selected running as our first sports activity for the following reasons: (a) Prevalence: Running is one of the most prevalent sports activities, does not require costly equipment, and due to the number of runners it would be easy for us to find participants. (b) Context variety: Unlike other activities such as skating, skiing, or swimming, which require a specific court/pool, running can be performed almost anywhere. As such, we may be able to collect an extensive range of contexts outside of indoor and outdoor courts, such as urban areas, forests, beaches, in the mountains, or on flat land---but also in varied weather and daylight conditions. 

We divided our data collection process into three phases (recording, cleaning up, and analysis).

For \textbf{recording}, we used either a GoPro HERO 10 or Ray-Ban Meta Smart Glasses, paired with a Garmin or Coros smartwatch. Participants (all of whom are authors on this publication) d first-person videos of their own typical activities over a 3-month period, including a variety of runs and races. We used a self-initiated think-aloud protocol where participants would turn on their recording device and record a brief spoken note along with contextualizing video whenever an information need occurred to them.    

For \textbf{cleaning up}, each individual reviewed their own audio and video and removed potentially-identifying or otherwise sensitive content, with a particular attention to other bystanders and other non-participant athletes. 
We then extracted transcripts from all videos for indexing and analysis. textbf{analysis}, we manually coded the extracted transcripts, identifying emergent themes for each individual and looking for patterns both within and across activities.

Each activity in the resulting corpus includes a \texttt{.fit} file containing biometrics for the run, a text transcript documenting information needs (stored as a \texttt{.txt} file), and a series of video snippets (as \texttt{.mp4} files).
Observations from our initial corpus (141 videos over 18 runs by 3 runners) show that information needs vary significantly between individuals, even when conducting similar activities. 
Over a three-month period, we examined our own information needs and found minimal overlap. Specifically, one of us identified 56 distinct information needs, another identified 37 distinct needs, and the third identified 42 distinct needs. 
Out of these, only 11 information needs were common across all three runners. 
These common needs included access to a route map, heart rate, speed, elevation, distance, elapsed time, water intake tracking, performance comparisons to other workouts, progress percentage, weather information, and pace planning.
\autoref{fig:teaser} shows examples of data need transcripts along with a screenshot of the video at the time that provides the context for the data needs.

\section{Work in Progress: Analysis Prototype}

\begin{figure}
    \centering
    \includegraphics[width=1.0\linewidth]{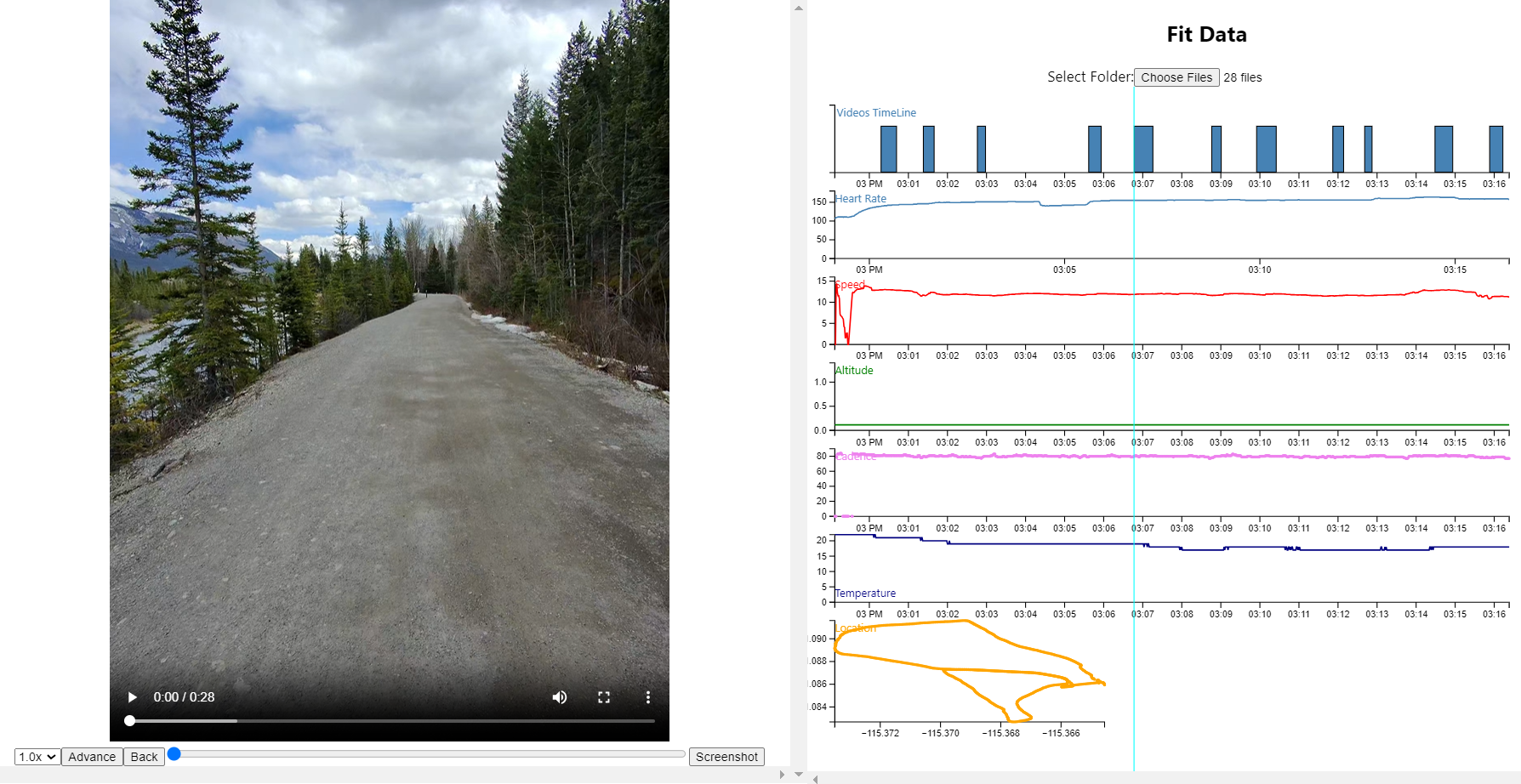}
    \caption{Our initial prototype interface showing video snippets and biometrics for a single activity.}
    \label{fig:prototype}
\end{figure}

\begin{figure}
    \centering
    \includegraphics[width=1.0\linewidth]{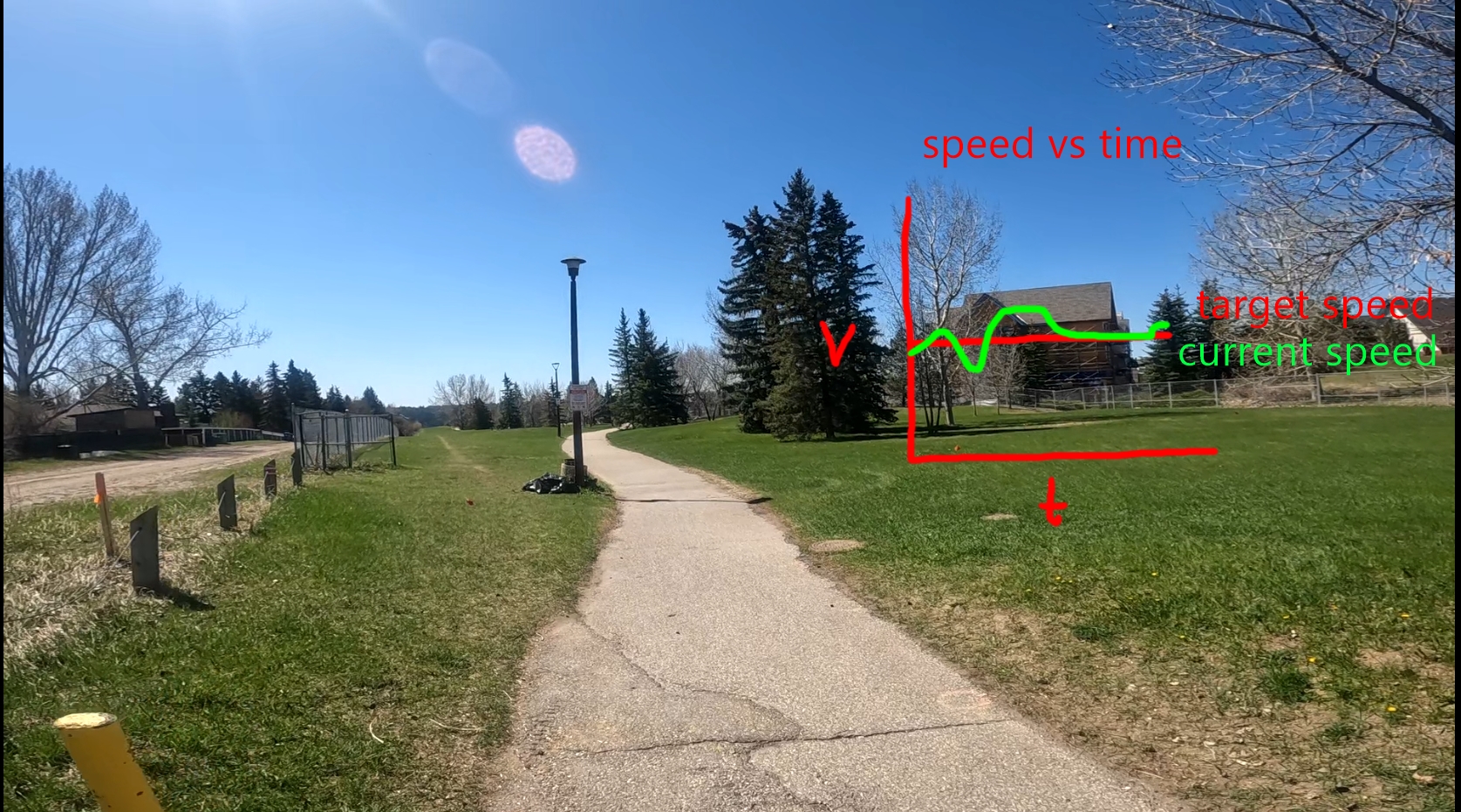}
    \caption{An early-stage sketch on top of a video snippet.}
    \label{fig:enter-label}
\end{figure}

To complement our process, we also initiated the development of a tool to help us analyze and aggregate first-person videos alongside the collected biometric data (\autoref{fig:prototype}). 
This tool allows us to view the video and corresponding biometric data in a synchronous manner, providing contextually-relevant information for each . 
Currently in its early phases, the tool allows to plot biometric data along with a timeline indicating when the videos were taken during the workout. %The biometric information is extracted from .fit files. 
We plan to continue refining and expanding its features to further support our research and development efforts.

Additionally, with this prototype, we created early-stage sketches on top of stills captured from these recorded videos. This approach helped us explore how these sketches can be overlaid on first-person points of view and what implications this might have for future tools and applications in similar contexts.

\section{Future Research: Opportunities \& Challenges}
Our initial data collection and analysis surfaced several recurring themes and opportunities related to information needs. These opportunities suggest potential approaches as we move forward with our prototype design. Given that this research project is in its early stages and our data collection corpus is limited, our findings are preliminary and require further validation. Nonetheless, these are the main opportunities we have identified so far:

\begin{itemize}
    \item \textbf{Variability Across Individuals:} Information needs can vary considerably across individuals and activity types, influenced by factors such as weather, time-of-day, mindset, and location. It remains to be seen how we can identify regular patterns even for individuals. Additionally, there is a need to explore whether some needs are highly specific to particular circumstances and mindsets; for instance, whether users require detailed guidance and feedback or prefer minimal, high-level data. How needs differ based on whether the activity is solitary or involves social or competitive elements also remains uncertain.

    \item \textbf{Variability Across Time:} Information needs can change considerably over time, even for the same individual. For example, a person training for a marathon may have different needs during the early stages of training compared to the final weeks before the event. Developing strategies and data references that adapt over time and context is an area that requires further investigation. Additionally, similar information needs from different participants might look different on the surface or vice versa, which complicates the process of identifying and addressing these needs effectively.

    \item \textbf{Relevance of Comparisons}: Many information needs involve comparisons to past data, other activities, or other individuals, raising the question of how to determine which comparisons are relevant given the vast range of possible comparisons. Information needs often extend beyond biometrics to include safety considerations (like the presence of bears), environmental conditions (weather conditions, trail closures), and personal preferences (such as music or historical context). 

    \item \textbf{Importance of Certain Needs and Avoiding Overload:} Some analyses may not be practical or necessary to perform in real-time during the activity. Identifying which analyses are meaningful and when they should be performed remains an open question. Additionally, there are times when individuals might prefer not to have access to data, focusing instead on connecting with nature or the activity itself. Balancing the provision of useful data with the option for minimal or no data is another area that needs careful consideration.

    \item \textbf{Influence of Data Collection on Data Needs:} The process of collecting data can influence the data needs, prompting new ideas and needs that might not have been considered otherwise. Managing this self-nurturing and iterative effect without complicating the design process is a valuable opportunity for innovation.

    \item \textbf{Broadening the Corpus:} The extensive data collection requirements, while crucial for detailed insights, limited participant diversity and may affect the generalizability of the findings. Moving forward, one potential avenue worth exploring could be leveraging accessible platforms like Strava for data collection. For example, a Strava user might store a recorded running workout privately for later processing. Although Strava only captures biometric data and would need to be aided by a pipeline for think-aloud information needs, there remains significant potential in finding ways to simplifying the data collection process and integrating less intrusive methods which could help in ensuring a wider range of participants.

\end{itemize}

These opportunities highlight the potential for designing tools that address varied and evolving information needs during physical activities. As we continue our research, these insights will inform our approach to developing effective and adaptable solutions.

\acknowledgments{
This work was partially supported by ANR-19-CE33-001, NSERC Discovery Grants [RGPIN-2021-02492], Alberta Innovates, the Canada Research Chairs program, and a University of Calgary PURE Award.}

\bibliographystyle{abbrv-doi}

\bibliography{Ahmed}
\end{document}